\documentclass[numreferences]{kluwer}

\usepackage[dvips]{epsfig}

\begin{document}

\begin{opening}

\title{\bf{Combinatorial Physics, Normal Order and Model Feynman Graphs}}
\author{Allan I. \surname{Solomon}$^{\dag\S}$}
\author{Pawel \surname{Blasiak}$^{\dag\ddag}$}
\author{Gerard  \surname{Duchamp}$^{\S\S}$}
\author{Andrzej \surname{Horzela}$^{\ddag}$}
\author{Karol A. \surname{Penson}$^{\dag}$}
\institute{\vspace{10pt}$^{\dag}$Universit\'{e} Pierre et Marie Curie\\
Laboratoire   de  Physique   Th\'{e}orique  des  Liquides, CNRS UMR 7600\\
Tour 16, $5^{i\grave{e}me}$ \'{e}tage, 4, place Jussieu, F 75252
Paris, Cedex 05, France\\
e-mail: penson@lptl.jussieu.fr\\\vspace{5pt}
$^{\ddag}$H. Niewodnicza{\'n}ski Institute of Nuclear Physics, Polish
Academy of Sciences\\
Department of Theoretical Physics\\
ul. Radzikowskiego 152, PL 31-342 Krak{\'o}w, Poland
\\e-mail: pawel.blasiak@ifj.edu.pl, andrzej.horzela@ifj.edu.pl \\\vspace{5pt}
$^\S$The Open University\\
 Physics and Astronomy Department\\
Milton Keynes MK7 6AA\\
e-mail: a.i.solomon@open.ac.uk\\\vspace{5pt}
$^{\S\S}$LIFAR, Universit\'{e} de Rouen\\
76821 Mont-Saint Aignan Cedex,
France\\
e-mail: gduchamp2@free.fr}
\begin{abstract}
The general normal ordering problem for boson strings is a combinatorial problem. In this talk we restrict ourselves to single-mode boson monomials. This problem leads to elegant generalisations of well-known combinatorial numbers, such as Bell and Stirling numbers. We explicitly give the generating functions for some classes of these numbers. Finally we show that a graphical representation of these combinatorial numbers leads to sets of model field theories, for which the graphs may be interpreted as Feynman diagrams corresponding to the bosons of the theory. The generating functions are the generators of the classes of Feynman diagrams.
\end{abstract}

\runningtitle{Combinatorial Physics, Normal Order and Model Feynman Graphs}

\runningauthor{Solomon, Blasiak, Duchamp, Horzela, Penson}


\end{opening}
\section{Boson Normal Ordering}
In this note we give a brief review of the combinatorial properties associated with the normal ordering of bosons, and the model Feynman graphs which result.

Combinatorial sequences appear naturally  in
the solution of the boson normal ordering problem \cite{BPS1}, \cite{AoC}. 

The normal ordering problem for canonical bosons  $[a,a^\dag]=1$ is
related to certain combinatorial numbers $S(n,k)$ called Stirling
numbers of the second kind through \cite{Katriel1}
\begin{eqnarray}
(a^\dag a)^n=\sum_{k=1}^nS(n,k) (a^\dag)^k a^k,
\end{eqnarray}
with corresponding numbers $B(n)=\sum_{k=1}^n S(n,k)$ called Bell
numbers. In fact, for physicists, these equations may be taken as the
{\em definitions} of the Stirling and Bell numbers. For quons
(q-bosons) satisfying $[a,a^\dag]_q\equiv aa^\dag-qa^\dag a=1$  a
natural q-generalisation \cite{Katriel2} of these numbers is
\begin{eqnarray}
(a^\dag a)^n=\sum_{k=1}^nS_{q}(n,k) (a^\dag)^k a^k.
\end{eqnarray}
In the canonical boson case, for integers
$n,r,s>0$ we define generalized Stirling numbers of the second
kind $S_{r,s}(n,k)$ through ($r\geq s$):
\begin{eqnarray}\label{C}
[(a^\dag)^ra^s]^n=(a^\dag)^{n(r-s)}\sum_{k=s}^{ns}S_{r,s}(n,k)(a^\dag)^ka^k,
\end{eqnarray}
as well as generalized Bell numbers $B_{r,s}(n)$ 
\begin{eqnarray}\label{D}
B_{r,s}(n)=\sum_{k=s}^{ns}S_{r,s}(n,k).
\end{eqnarray}
For both $S_{r,s}(n,k)$ and $B_{r,s}(n)$ exact and explicit formulas
have been found \cite{BPS1, AoC}.  We refer the interested reader to these sources for further information on those extensions.  However, in this note we shall only deal with the classical Bell and Stirling numbers, corresponding to $B_{1,1}(n)$ and $S_{1,1}(n)$ in our notation.  
\section{Generating Functions}
In general, for combinatorial numbers $g(n)$ we may define an {\em
exponential generating function} $G(x)$ through \cite{Wilf}
\begin{equation}
G(x) = \sum_{n=0}^{\infty} g(n)\frac{x^n}{n!}.
\end{equation}
For the Bell numbers, the generating function takes the particularly nice form\cite{Comtet}
\begin{equation}
G(x) = \sum_{n=0}^{\infty} B(n)\frac{x^n}{n!}=\exp(\exp(x)-1).
\end{equation}
Some initial terms of the sequence $\{B(n)\}$ are $\{1,2,5,15,52,203,877,\ldots \}$.
\section{Graphs}
A convenient way of representing combinatorial numbers is by means of {\em graphs}.  To illustrate this, we now consider a graphical method for illustrating the combinatorial numbers associated with the normal order expansion of $(a^{\dagger}a)^n$.
\begin{figure}
\vspace{1cm}
\begin{center}\resizebox{10cm}{!}{\includegraphics{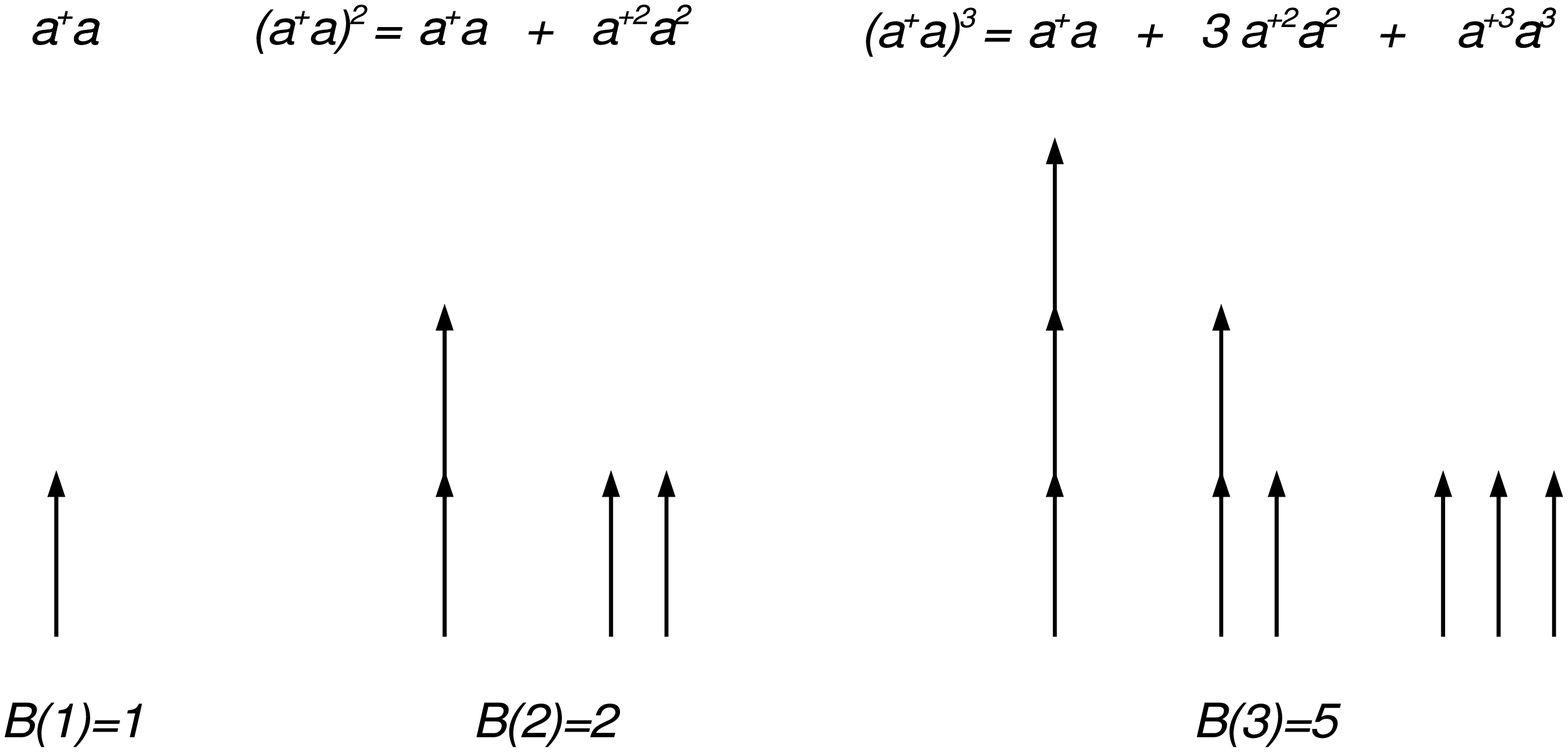}}
\caption{Arrow graphs for $(a^{\dagger}a)^n \; \; \; n=1,2,3.$}\label{FigA}
\end{center}
\end{figure}

A single arrow represents a ``time-segment'' of a line corresponding to the  ``propagator'' $(a^{\dagger}a)$.  Thus we may concatenate one, two or more arrows to form a single line, or propagator $(a^{\dagger}a)$.  However, two lines correspond to two distinct propagators ${a^{\dagger}}^2 a^2$, and so on.  Further, the constitutent arrows are labelled, for example  by time, and so they may only be concatenated respecting the time ordering.  These rules are illustrated by the  diagrams of Figure 1, in which we consider the cases of 1, 2,  and 3 arrows respectively. We have pre-emptively labelled these numbers as Bell numbers - which fact we demonstrate below. For the case of 4 arrows (Figure 2) we have additionally given the individual associated symmetry factors (in fact Stirling numbers) which add to $B(4)=15$.  It should be clear from this illustration how the time-ordering rules are applied to give the symmetry coefficients.  Thus there is only one way in which we can concatenate 4 arrows respecting time-ordering (first grouping), 4 ways in which we can divide the arrows into a set of 3 and 1, and so on. 

\begin{figure}
\vspace{1cm}
\begin{center}\resizebox{10cm}{!}{\includegraphics{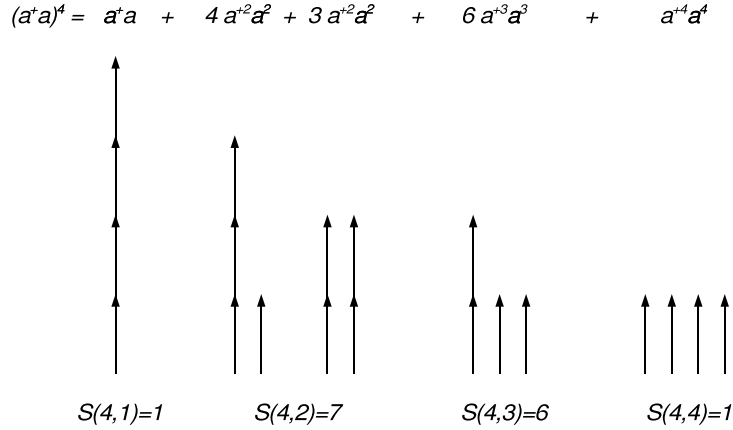}}
\caption{Arrow graphs for $(a^{\dagger}a)^4.$}\label{FigB}
\end{center}
\end{figure}

From these first few examples it would seem that these graphs are essentially like the Feynman Diagrams of a zero-dimensional (no integration) Model Field Theory associated with 
$H =a^{\dagger}a$.  In other words, at order $n$ the total number of graphs would appear to be $B(n)$ while the individual coefficients of  
${a^{\dagger}}^{k}a^k$ are given by $S(n,k)$.  In order to show that this is indeed the case, we must be able to count the number of graphs 
associated with a given number $n$ of arrows. 
To do this we can use the First of Three Great Results.

\section{First Great Result}
This First Great Result is sometimes known as the {\em  connected graph
theorem} \cite{cgt}.  It states that if $C(x)=\sum_{n=1}^{\infty} c(n)x^n/n!$ is the generating function of {\em labelled connected} graphs, {\em viz.} $c(n)$ counts the number of connected graphs of order $n$, then 
\begin{equation}
A(x) = \exp(C(x))
\end{equation}
is the generating function for {\em all} graphs.

We may apply this very simply to the case of the arrow graphs above. For each order $n$, the connected graphs consist of the single graph obtained by concatenating all the arrows into one propagator. Therefore for each $n$ we have $c(n)=1$; whence, $C(x)=exp(x)-1.$  It follows that the generating function for all the arrow graphs $A(x)$ is given by
\begin{equation}
A(x)=\exp(\exp(x)-1)
\label{bell}
\end{equation}
which is the generating function for the Bell numbers.

Such graphs  may be generalised to give  graphical representations for
the extensions $B_{r,s}(n)$ \cite{tobepub}.

However, just as an abstract group is capable of more than one
presentation, there are many graphical representations for a given
combinatorial sequence; and we now give an alternative one for the
numbers $B(n), S(n,k)$ due to Bender and collaborators \cite{ben1},
\cite{ben2}.
\section{Second Great Result}
As before, we shall be counting lines.  A line starts from a white dot, the {\em origin}, and ends at a black dot, the {\em vertex}. What we refer to as {\em origin} and {\em vertex} is, of course, arbitrary. At this point there are no other rules, although we are at liberty to impose further restrictions; a white dot may be the origin of 1,2,3,... lines, and a black dot the vertex for 1,2,3,... lines. We may further associate {\em strengths} $V_s$ with each vertex receiving $s$ lines, and  multipliers $L_m$ with a white dot which is the origin of $m$ lines.   Again  $\{V_s\}$ and $\{L_m\}$ play symmetric roles; in this note we shall only consider cases where the $L_m$ are either $0$ or $1$. 

We illustrate these rules for four different graphs corresponding to $n=4$.
\begin{figure}
\vspace{1cm}
\begin{center}\resizebox{10cm}{!}{\includegraphics{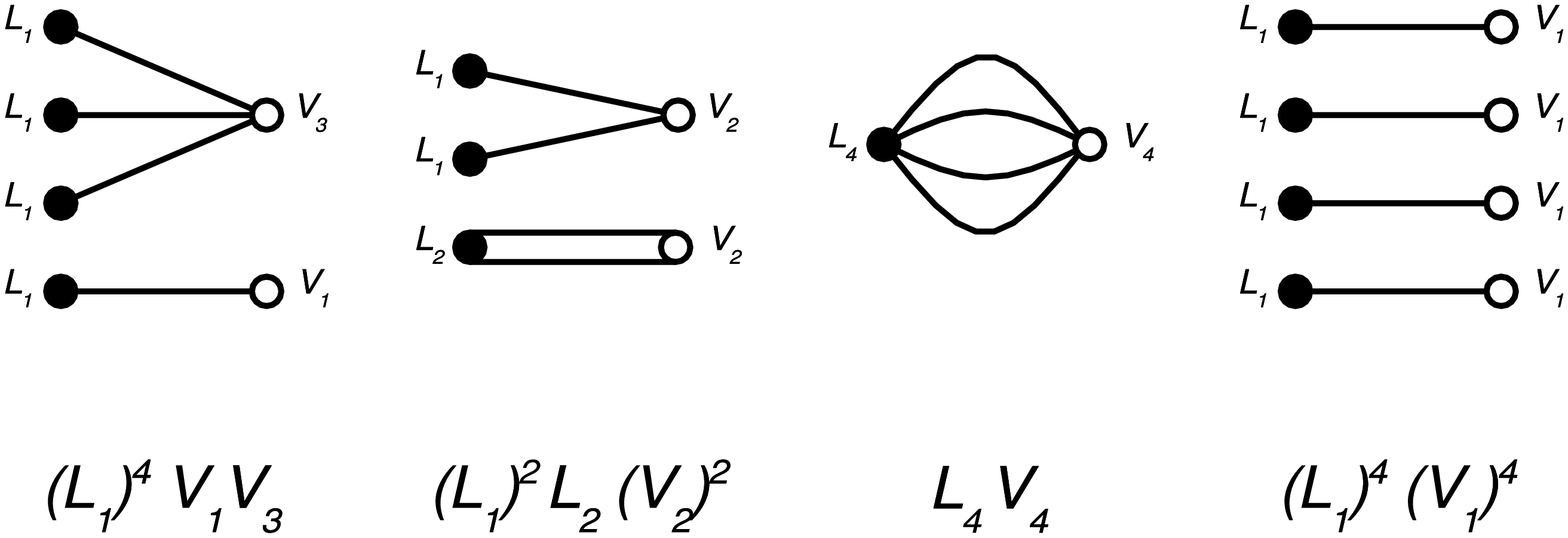}}
\caption{Some examples of 4-line graphs.}\label{FigC}
\end{center}
\end{figure}

There is an generating function $G(x,V,L)$ which counts the number
$g(n)$ of graphs with $n$ lines arising from the above rules \cite{Vasiliev}:
\begin{eqnarray}
G(x,V,L)&=&\left.\exp(\sum_{m=1}^{\infty} L_m \frac{x^m}{m!}\frac{d^m}{dy^m}) \exp(\sum_{s=1}^{\infty} V_s \frac{y^s}{s!})\right|_{y=0} \nonumber\\
        &\equiv&\sum_{n=0}^\infty g(n)\frac{x^n}{n!} \label{fgr}
\end{eqnarray}

Consider the following example: $L_m=0$ for all $m \neq1$, that is, we allow only one line from each origin (with multiplier 1); there is no restriction on the number of lines to a vertex, and $V_s =1$ for all $s$.  We give an example of the cases $n=1,2,3,4$ in Figure 4.  Note that for correct counting the lines should be labelled, as they were in the case of the arrows above. 

The generating function $G(x)$ which counts the lines corresponding to the above  rules follows immediately from Eq.(\ref{fgr}) 
\begin{eqnarray}
G(x)&=&\left.\exp(\sum_{m=1}^\infty L_m \frac{x^m}{m!}\frac{d^m}{dy^m})
\exp(\sum_{s=1}^\infty  V_s \frac{y^s}{s!})\right|_{y=0} \nonumber \\
        &=& \left.\exp( \frac{x}{1!}\frac{d}{dy}) \exp(\sum_{s=1}^\infty  \frac{y^s}{s!})\right|_{y=0}\nonumber \\
        &=& \left.\exp(x d/dy) \exp(e^y -1)\right|_{y=0}\nonumber \\
       &=& \exp(e^x -1) \equiv \sum_{n=0}^\infty {B(n) \frac{x^n}{n!}}.
\end{eqnarray}

The penultimate step is a consequence of the Taylor expansion.

We thus have yet another representation of the integer sequence
$\{B(n)\}$.  Note that when $L\equiv \{L_m\}$ and $V \equiv \{V_s\}$ are integer
sequences  we obtain an integer sequence from Eq.(\ref{fgr}).  A
convenient method of obtaining the resulting integer sequence is
afforded by the next useful result.
\begin{figure}
\vspace{1cm}
\begin{center}\resizebox{10cm}{!}{\includegraphics{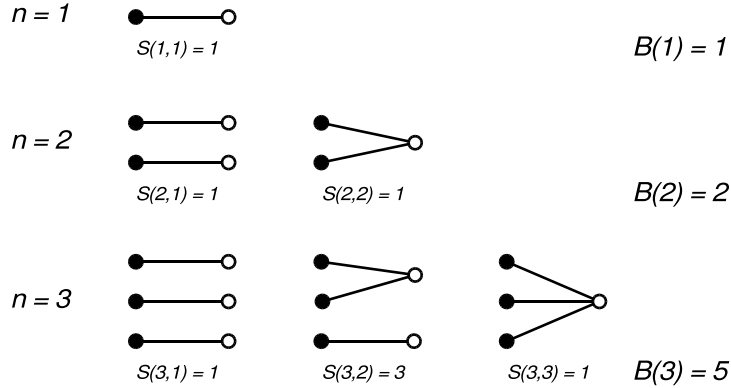}}
\caption{Graphs of second type for $B(n),\; \; \; n=1,2,3.$}
\end{center}
\end{figure}

\section{Third Great Result}
Straightforward manipulation of series shows the following:
Define
\begin{equation}
A_i(x) = \sum_{n=0}^{\infty} a_i(n) \frac{x^n}{n!} \; \; \; \; \; i=1,2,3.
\label{genprod}
\end{equation}
Then if
\begin{equation}
\left.A_1(x d/dy) A_2(y)\right|_{y=0} = A_3(x)
\end{equation}
we have
\begin{equation}
a_3(n)=a_1(n) a_2(n).
\label{prod}
\end{equation}
This is a useful and rather surprising equality.
 
Using the results of the previous section, Eq.(\ref{prod}) enables us
to create graphical representations of {\em products} of integral
sequences.  For example:  if in the case above  we chose $L_m=1$ for
all $m$, enabling any number of  lines from each origin (with
multiplier 1), the resulting sequence of graphs would have given us a
representation of the integer sequence $\{B(n)^2\}$ \cite{ben1}.

We exemplify the possibilities offered by application of the Third Great Result by our two final examples.

\underline{Example 1:} Generating function for the sequence
$\{B(n) B(n+1)\}$.

From Eqs.(\ref{genprod}) and (\ref{prod}), the required generating function is given by
\begin{equation}
G(x,V,L) = \left.A_1(x d/dy) A_2(y)\right|_{y=0}
\end{equation}
where $A_1(x)$ is the generating function for $\{B(n)\}$ and $A_2(x)$ is that of $\{B(n+1)\}$. Note that $A_1(x)=\exp(\exp(x)-1)$ from Eq.(\ref{bell}), while $A_2(x)=(d/dx)A_1(x)=\exp(\exp(x)-1+x)$. 

This shows that the graphs for the sequence $\{B(n)B(n+1)\}$ may be
obtained by putting $L_m=1$ for all $m$, so that
there are any number of lines emanating from an origin, and with
multiplicity 1. For the sequence $\{B(n+1)\}$ we have $V_s=1$ for all
$s\neq2$ and $V_2=2$, so that any number of lines may end at a vertex, and all have strength 1 except for the case where {\em two} lines meet at a vertex, when the strength is 2.
The generating function
is, according to Eq.(\ref{prod}), 
\begin{equation}
G_1(x,V,L) = \sum_{n=0}^{\infty} B(n)B(n+1) \frac{x^n}{n!}.
\end{equation}
It may be explicitly obtained after  some formal algebraic
manipulation based on the Dobi\'nski formula \cite{Dob}
\begin{equation}
B(n) = \frac{1}{e} \sum_{k=0}^{\infty} \frac{k^n}{k!}
\end{equation}
as
\begin{equation}
G_1(x,V,L) = \sum_{k=0}^{\infty} \frac{1}{k!}\exp(\exp((k+1)x)-2).
\label{g1}
\end{equation}
The formal series (\ref{g1}) diverges for all $x>0$, although having finite Taylor coefficients for $x=0$. Such formal series are nevertheless useful in representing integer sequences. 

\underline{Example 2:} In our last example we retain all the derivative terms in Eq.(\ref{genprod}) but choose $V_1=V_2=1; V_s=0, s>2$, thus allowing vertices where at most {\em two} lines meet. 
The corresponding generating function is defined by
\begin{equation}
G_2(x,V,L) = \sum_{n=0}^{\infty} B(n)I(n) \frac{x^n}{n!}
\end{equation}
where the {\em Involution numbers} $I(n) = 1,2,4,10,26,76,\ldots$ are defined through their generating function
\begin{equation}
G_I(x) = \exp(x+\frac{x^2}{2}) = \sum_{n=0}^{\infty} I(n) \frac{x^n}{n!}
\end{equation}
and are special values of the Hermite polynomials $H_n(x)$
$$ I(n) = H_n(\frac{1}{\sqrt{2} i})/(-\sqrt{2} i)^n . $$
Consequently
\begin{eqnarray}
G_2(x,V,L) &=& \sum_{n=0}^{\infty} B(n) H_n(\frac{1}{\sqrt{2} i})
(\frac{-x}{\sqrt{2} i})^n/n! \label{g2}\\
&=&\sum_{k=0}^\infty \frac{1}{k!}\exp(kx(1+\frac{kx}{2})-1).
\label{e2} \end{eqnarray}
In obtaining Eq.(\ref{e2}) we have used the standard form of the
generating function of the Hermite polynomials.
Again, we must consider the series Eqs.(\ref{g2}) and (\ref{e2}) as
formal power series, since for example they diverge for $x>0$. 

In conclusion, we emphasize that the expressions of Eqs.(\ref{g1}) and (\ref{e2}) constitute exact solutions of Model Field Theories defined by the appropriate sets $\{V_s\}$ and $\{L_m\}$.  Many other applications and extensions of the ideas sketched in this note will be found in \cite{tobepub}.
\begin{acknowledgements}

We thank  Carl Bender and  Itzhak Bars for interesting  discussions.

\end{acknowledgements}

\end{document}